\documentclass[aps,prl,preprint,groupedaddress]{revtex4}
\usepackage{graphicx}
\begin{document}

\title{Rippled Cosmological Dark Matter\\
from Damped Oscillating Newton Constant}

\author{Aharon Davidson}
\email[Email:~]{davidson@bgumail.bgu.ac.il}
\homepage[HomePage:~]{www.bgu.ac.il/~davidson}
\affiliation{Physics Department, Ben-Gurion University, Beer-Sheva
84105, Israel \\
{\tt davidson@bgumail.bgu.ac.il}}

\date{June 15, 2004}

\begin{abstract}
	Let the reciprocal Newton 'constant' be an apparently
	non-dynamical Brans-Dicke scalar field damped oscillating
	towards its General Relativistic VEV.
	We show, without introducing additional matter fields
	or dust, that the corresponding cosmological evolution
	averagely resembles, in the Jordan frame, the familiar
	dark radiation $\rightarrow$ dark matter $\rightarrow$
	dark energy domination sequence.
	The fingerprints of our theory are fine ripples, hopefully
	testable, in the FRW scale factor; they die away at the
	General Relativity limit.
	The possibility that the Brans-Dicke scalar also serves
	as the inflaton is favorably examined.
\end{abstract}
 
\pacs{98.80.-k, 95.35.+d, 04.50.+h}
\maketitle

$\Lambda$CDM is perhaps the best cosmological model available.
Only two free parameters, controlling the density of dark matter\cite{DM}
and dark energy\cite{DE} (in the form of a cosmological constant
$\Lambda$) suffice to parametrize the apparent clash
between general relativity (GR) and contemporary cosmology.
The notion of dark matter is commonly used, but so far we do not
have any clue whether dark matter is a fact (in which case identifying
its constituents is a major particle physics challenge) or else an artifact
(effectively induced when casting the field equations of some beyond
GR theory into the traditional Einstein equations format).
In this Letter we show, without introducing additional matter fields
or dust, that the general relativistic cosmological interpretation of
a damped oscillating\cite{damp} Newton constant, sliding toward
its constant VEV, is of a dark radiation $\rightarrow$ dark matter
$\rightarrow$ dark energy dominated Universe.
The cosmological fingerprints of such a theory are fine (suppressed
at the GR limit) ripples, hopefully testable, in the FRW scale factor.

\medskip
The general idea that Nature's fundamental constants may vary in
space and time is not new.
The theoretical motivation for a varying Newton constant, usually
accompanied by a varying fine structure constant, ranges from
Kaluza-Klein theory\cite{KK}, Brans-Dicke theory\cite{BD} and
dilaton/conformal\cite{CG} gravity all the way to brane
models\cite{brane} and even super-string/M-theory\cite{String}
cosmology.
One may add to this list k-essence models\cite{kessence}, and in
particular, Zee's broken-symmetric model\cite{Zee} of gravity. 
From the experimental\cite{review} side, however, there is so far
no indication to support a varying Newton constant, with the major
cosmological constraints coming from CMB\cite{CMB} and
Nucleosynthesis\cite{NS}.

\medskip
Let our starting point be the Brans-Dicke action
\begin{equation}
	{\cal S}=-\frac{1}{16\pi}\int \left(\phi {\cal{R}}
	+\frac{\omega_{BD}}{\phi}
	g^{\mu\nu}\partial_{\mu}\phi\partial_{\nu}\phi
	+V(\phi)+{\cal L}_{m}\right)
	\sqrt{-g}~d^{4}x  ~,
\end{equation}
with $\phi(x)^{-1}$ locally representing the varying Newton
'constant'.
The main role of the scalar potential $V(\phi)$ will be
to allow, by virtue of Zee mechanism\cite{Zee}, for the
spontaneous emergence of General Relativity as the BD
scalar approaches its constant VEV.
In the Jordan frame used, the matter Lagrangian
${\cal L}_{m}$ is minimally coupled to gravity, and does
not couple to $\phi(x)$.
In turn, all matter field equations exhibit the standard
general relativistic structure, and test particles fall freely
along geodesics of the (thereby physical) metric $g_{\mu\nu}$.
Note that a scale transformation
$g_{\mu\nu}\rightarrow \phi^{-1}g_{\mu\nu}$, which is
definitely not a general coordinate transformation, would
give rise to a proper-sign scalar kinetic term plus a non-trivial
scalar-matter coupling in the so-called Einstein frame.
To focus on the physics beyond-GR, however, let us
momentarily switch off the matter Lagrangian ${\cal L}_{m}$.
We will eventually have to switch it on, e.g. by adding
a tiny amount of ordinary dust, as otherwise it will be
impossible to tell which metric is actually used by
physical rods and clocks.

\medskip
On some effective potential grounds (to be clarified soon),
we choose the special case $\omega_{BD}=0$.
Counter intuitively, this will not make $\phi(x)$ non-dynamical,
and furthermore, would not contradict the well-tested GR in
the solar system provided\cite{w} the potential is sufficiently
confining to render the BD scalar essentially constant.
Although the conventional scalar kinetic term $g^{\mu\nu}
\phi_{;\mu}\phi_{;\nu}$ has been unconventionally waived
away, second order derivatives of $\phi$ will still enter the
game, owing to the fact that
$\sqrt{-g}\phi g^{\mu\nu}\delta{\cal R}_{\mu\nu}$
is no longer a total divergence.
Associated with the above action principle, with
$\omega_{BD}=0$ enforced and in the momentarily absence
of ${\cal L}_{m}$, are the simplified field equations 
\begin{eqnarray}
	&\phi ({\cal R}_{\mu\nu}
	-\textstyle{\frac{1}{2}}g_{\mu\nu}{\cal R})=
	-\phi_{;\mu\nu}
	+g_{\mu\nu}g^{\alpha\beta}\phi_{;\alpha\beta}+
	\textstyle{\frac{1}{2}}g_{\mu\nu}V(\phi) ~,&\\
	&\displaystyle{\frac{\partial V(\phi)}{\partial\phi}+
	{\cal R}=0} ~.&
\end{eqnarray}	
By substituting ${\cal R}$ into the trace of the gravitational field
equation, the familiar Klein-Gordon equation, governed by an
effective potential $V_{eff}(\phi)$, makes its appearance
\begin{equation}
	g^{\mu\nu}\phi_{;\mu\nu}=
	\frac{1}{3}
	(\phi \frac{\partial V}{\partial \phi}-2V)
	\equiv 
	\frac{\partial V_{eff}(\phi)}{\partial \phi}  ~.
	\label{Veff}
\end{equation}
The emergence of such an effective potential actually characterizes
the $\omega_{BD}= 0$ case.
The most general quadratic potential, for example,
\begin{equation}
	V(\phi)=\alpha+2\beta\phi+\gamma\phi^{2} ~,
	\label{V}
\end{equation}
gets translated into
\begin{equation}
	V_{eff}(\phi)= -\frac{1}{3}(2\alpha\phi+\beta\phi^{2})
	+const ~.
\end{equation}
We shall see that, for a suitable range of parameters, such a
potential can deliver stable attractive gravity, meaning
\begin{equation}
	\langle\phi\rangle \equiv G^{-1}>0 ~.
\end{equation}

Within the framework of FRW cosmology, with its spatially flat
$k=0$ variant favored on simplicity (as well as experimental)
grounds, the two independent field equations take the form
\begin{equation}
	\dot{\phi}H+\phi H^2=
	\displaystyle{\frac{1}{6}V(\phi)} ~,\quad
	\dot{H}+2H^2 =
	\displaystyle{\frac{1}{6}\frac{\partial V(\phi)}{\partial\phi}} ~,
	\label{FieldEqs}
\end{equation}
with $H=\dot{a}/a$ serving as the Hubble constant.
The gravitational field equations can be recasted into the
Einstein format, that is
${\cal R}_{\mu\nu}-\textstyle{\frac{1}{2}}g_{\mu\nu}{\cal R}=
{\cal T}_{\mu\nu}^{GR}$, thereby defining an effective (from
GR point of view) conserved energy/momentum tensor.
The associated effective energy density and pressure are given
respectively by
\begin{equation}
	\rho^{GR} = \displaystyle{-3H\frac{\dot{\phi}}{\phi}
	+\frac{1}{2\phi}V }~,~~
	P^{GR}= \displaystyle{\frac{\ddot{\phi}}{\phi}
	+2H\frac{\dot{\phi}}{\phi}-\frac{1}{2\phi}V} ~.
\end{equation}

The GR limit of our generalized Brans-Dicke theory, if exists, can
be straight forwardly derived, and for the quadratic potential
eq.(\ref{V}), one finds
\begin{equation}
	\phi(t) \rightarrow -\frac{\alpha}{\beta}
	\equiv G^{-1},~
	H^{2}(t) \rightarrow \frac{1}{6}
	\left(\beta-\frac{\alpha\gamma}{\beta}\right)
	\equiv \frac{\Lambda}{3} ~.
	\label{VEV}
\end{equation} 
Clearly, to assure the attractive nature of gravity, one must require
$\alpha/\beta<0$.
This, together with the fact that a non-negative $H^{2}$ implies
$\beta\geq \alpha\gamma/\beta$, furthermore tells us that
$\gamma\geq\beta^{2}/\alpha$.
We can finally close the allowed range of parameters, namely
\begin{equation}
	\alpha>0 ~,~ \beta<0 ~,~
	\gamma\geq\frac{\beta^{2}}{\alpha}>0 ~,
\end{equation}
by insisting on having $V_{eff}(\phi)$ bounded from below.
It is important to notice, however, owing to the fact that the effective
potential is $\gamma$-independent, that the background cosmological
constant $\Lambda \geq 0$ can be arbitrarily fine tuned without
affecting $V_{eff}(\phi)$.

\begin{figure}[ht]
	\includegraphics[scale=0.27]{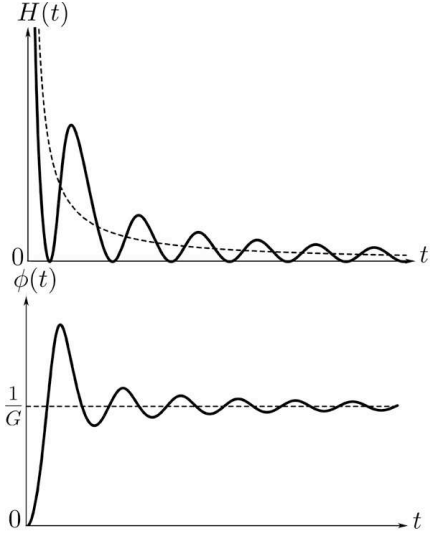}
	\caption{\label{Hphi1} The Hubble constant $H(t)$ and
	the reciprocal Newton constant $\phi(t)$ evolution for
	$\Lambda=0$. Time averaging over the ripples, justified
	for large $\omega t$, resembles a General Relativistic
	matter dominated Universe (dashed).}
\end{figure}

\medskip
On pedagogical grounds, we first discuss the case of a flat Lorentzian
background, and then generalize the scope to allow for the cosmological
deSitter background.
The constraint $\Lambda=0$ can be conveniently parameterized by
$\alpha=6\omega^{2}G^{-1}$, $\beta=-6\omega^{2}$, and
$\gamma=6\omega^{2}G$ (this mass scale $\omega$ has nothing
to do with the Brans-Dicke parameter $\omega_{BD}$), so that
\begin{equation}
	V(\phi)=\frac{6\omega^2}{G}(1-G\phi)^2
	~ \Rightarrow ~
	V_{eff}(\phi)=\frac{1}{3G}V(\phi)~.
\end{equation}
Substituting one field equation
\begin{equation}
	G\phi(t)=1+\frac{1}{2\omega^{2}}
	\left(\dot{H}+2H^{2}\right)
	\label{phi1}
\end{equation}
into the other, see eqs.(\ref{FieldEqs}), leaves us with the
second order master equation for the Hubble constant	
\begin{equation}
	H\ddot{H}+3H^{2}\dot{H}-\frac{1}{2}\dot{H}^{2}
	+2\omega^{2}H^{2}=0 ~,
	\label{H1}
\end{equation}
with $\omega$ being the only mass scale involved.
The numerical solution of the latter equations is plotted in
Fig.\ref{Hphi1}.
%%%%%%%  Fig.1
	
\medskip
\noindent $\bullet$
At early times ($\omega t \ll 1$), the solution of eq.(\ref{H1})
is characterized by one relevant dimensionless constant of
integration $p$
\begin{eqnarray}
	H(t) &\simeq& \frac{1}{2t}
	+\frac{4\omega p}{5}(\omega t)^{1/2} ~,
	\label{HBB}\\
	G\phi(t) &\simeq&
	\frac{p}{(\omega t)^{1/2}}+\frac{4p^{2}}{5}\omega t ~.
	\label{phiBB}
\end{eqnarray}
Eq.(\ref{HBB}) tells us that, at early times, the metric associated
with a varying Newton constant reminds us of a \emph{radiation
dominated} Universe recovering from a singular Big-Bang.
Depending on the sign of $p$, weak initial gravity may be
attractive $(p>0)$ or else repulsive $(p<0)$.
It is only the special $p=0$ case for which the $\phi$-expansion
starts with
\begin{equation}
	G\phi(t)\simeq \frac{4}{5}\omega^{2}t^{2} ~,
	\label{hot}
\end{equation}
describing strong (small $\phi$), necessarily attractive (positive
$\phi$), initial gravity.
The latter case of preference can be singled out on finiteness
grounds when noticing that, as $t \rightarrow 0$, the Ricci scalar
and the Lagrangian behave like ${\cal{R}} \simeq 12\omega^{2}$
(rather than $\sim p t^{-1/2}$) and
$\phi {\cal R}+V \simeq 6\omega^{2}G^{-1}$
(rather than $\sim p^{2}t^{-1}$), respectively.

\begin{figure}[ht]
	\includegraphics[scale=0.27]{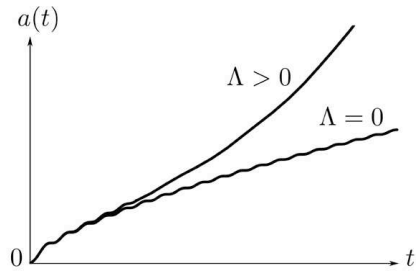}
	\caption{\label{a} Rippled FRW scale factor: Dark
	matter ripples are only $1/t$-suppressed, whereas
	dark energy ripples are exponentially suppressed. The
	ripples die away at the $\omega t\rightarrow \infty$
	general relativity limit.}
\end{figure}

\medskip	
\noindent $\bullet$
At later times ($\omega t \gg 1$), asymptotically approaching
GR, the solution exhibits a particular $1/t$-behavior modulated
by damped oscillations.
This can very neatly numerically verified by plotting $H^{-1}(t)$
and noticing that all its minima (corresponding to the maxima of
$H(t)$) lie on a straight line.
Quite serendipitously, however, is the fact that the slope of this
line is $\frac{3}{4}$.
This observation is fully captured by the analytic expansion
\begin{eqnarray}
	&\displaystyle{H(t)\simeq
	\frac{4\cos^{2}{\omega t}}{3t}
	\left(1-\frac{\sin{2\omega t}
	+\frac{5}{4}\tan{\omega t}
	+\widetilde{p}}{2\omega t}\right)} ~,&
	\label{Hlate}\\
	 & \displaystyle{G\phi(t) \simeq
	 1-\frac{2\sin{2\omega t}}{3\omega t}}
	 ~~ \Rightarrow ~~
	\frac{\dot{\phi}(t)}{\phi(t)}\simeq
	-\frac{4\cos{\omega t}}{3t} ~,
	 \label{philate}& 
\end{eqnarray}
where $\widetilde{p}$ is a constant of integration (presumably
related to the previous $p$).
The corresponding FRW scale factor $a(t)$ can be integrated
now to establish our main result
\begin{equation}
	\fbox{$\displaystyle{a(t)\sim t^{2/3}\left(
	1+\frac{\sin{2\omega t}
	+\widetilde{p}}{3\omega t}\right)}$}
	\label{FRW}
\end{equation}
The full numerical solution for $a(t)$ is plotted in Fig.\ref{a}
(focus at the moment on the $\Lambda=0$ branch).
As clearly demonstrated in this Figure, the FRW evolution highly
resembles a \emph{dark matter dominated} Universe.
Moreover, the non-trivial correlation
\begin{equation}
	a(t)\sim t^{2/3}~\leftrightarrow~
	\phi(t)\simeq G^{-1}
\end{equation}
leads to the pleasing conclusion that the emerging matter
dominated Universe is in fact general relativistic.
%%%%%%%  Fig.2

\begin{figure}[ht]
	\includegraphics[scale=0.27]{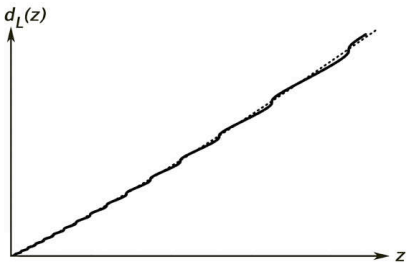}
	\caption{\label{Hubble} Hubble Plot: The luminosity distance
	$d_{L}(z)$, plotted as a function of the redshift $z$, is confronted
	with its corresponding matter-dominated GR rival (dashed).}
\end{figure}

\medskip
Two practical questions immediately arise:
(i) How large is this $\omega$ frequency? and
(ii) Can one experimentally probe the rippled structure?
Following the Nyquist-Shannon theorem\cite{sample}, we need an
experimental sampling frequency $\omega_{NS} \geq 2\omega$
in order to faithfully reconstruct the original signal and avoid aliasing.
Obviously, if $\omega^{-1}$ is many orders of magnitude smaller
than the age of the Universe, one would not be able to tell the
actual FRW scale $a(t)$ from its coarse grained average 
\begin{equation}
	\overline{a}(t) \equiv
	\frac{1}{2T}\int_{t-T}^{t+T}
	a(t^{\prime})dt^{\prime}
\end{equation}
for some $T\simeq$ a few $\omega^{-1}$.
Tiny deviations from the average would presumably be interpreted
as measurement errors.
In some sense, this reminds us of a tracking oscillating energy
model\cite{oscillating} where the potential takes the form of a
decaying exponential with small perturbations superimposed.
To appreciate the way the GR limit is approached, consider a
physical quantity which can be calculated by means of Eq.(\ref{FRW}),
e.g. the luminosity distance $d_{L}(z)$, and compare it with the
standard (GR) Hubble plot for a matter dominated Universe
(see Fig.\ref{Hubble}).
%%%%%%%  Fig.3
	
\begin{figure}[ht]
	\includegraphics[scale=0.27]{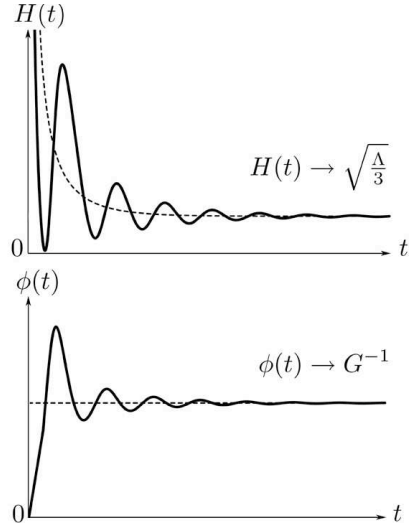}
	\caption{\label{Hphi2}The Hubble constant $H(t)$ and
	the reciprocal Newton constant $\phi(t)$ evolution for
	$\Lambda> 0$. $\Lambda$CDM-like evolution (dashed)
	is recovered by time averaging over  the ripples.}
\end{figure}

Attempting to be a bit more realistic, we now switch on the
background cosmological constant $\Lambda$.
This is done by shifting the potential $V_{\Lambda = 0}(\phi)$, namely
\begin{equation}
	V_{\Lambda \neq 0}(\phi) =
	V_{\Lambda = 0}(\phi) + 2G\Lambda \phi^{2} ~.
\end{equation}
Notice that $V_{eff}(\phi)$ is not sensitive to such a transformation.
In turn, $V(\phi)$ and $V_{eff}(\phi)$ do not share anymore the
same absolute minimum, and the vacuum of the theory, reflecting
the minimization of $V_{eff}(\phi)$, is associated now with
$\displaystyle{\langle {\cal R} \rangle=
-\langle \frac{\partial V}{\partial\phi}\rangle=-4\Lambda}$.
In some sense, this reminds us of Ref.\cite{R+1/R}, where the
constant curvature vacuum solution is deSitter rather than Minkowski.
At any rate, in the presence of $\Lambda$, Eq.(\ref{phi1}) is
generalized into
\begin{equation}
	G\phi(t)=1+\frac{\dot{H}+2H^{2}-\frac{2}{3}\Lambda}
	{2(\omega^{2}+\frac{1}{3}\Lambda)}
	\label{phi2}
\end{equation}
whereas eq.(\ref{H1}) acquires a non-homogeneous term
\begin{equation}
	H\ddot{H}+3H^{2}\dot{H}-\frac{1}{2}\dot{H}^{2}
	+2\omega^{2}H^{2}=\frac{2}{3}\omega^{2}\Lambda ~.
	\label{H2}
\end{equation}
The associated numerical solution is plotted in Fig.\ref{Hphi2}.
%%%%%%%  Fig.4

\medskip
\noindent $\bullet$
At early times ($\omega t \ll 1$), as expected, the $\Lambda=0$
solution eqs.(\ref{HBB},\ref{phiBB}) remains qualitatively the same.
The leading $\Lambda$-modified term in $H(t)$ is only
${\cal O}(\omega^{3}t^{3})$, while $\phi(t)$ is only stretched by
an overall factor $3\omega^{2}/(\Lambda+3\omega^{2})$.
As before, strong attractive initial gravity is encountered for the
$p=0$ choice, with eq.(\ref{hot}) staying intact.

\medskip
\noindent $\bullet$
At later times ($\omega t \gg 1$), however, the $\Lambda > 0$
solution is entirely different from the $\Lambda = 0$ solution
eqs.(\ref{Hlate},\ref{philate}).
To see what are the possibilities, take
$H(t)=\sqrt{\Lambda/3}+\delta(t)$ and linearize eq.(\ref{H2})
to read
\begin{equation}
	\ddot{\delta}+\sqrt{3\Lambda}\dot{\delta}
	+4\omega^{2}\delta=0 ~.
\end{equation}
So, it is either damped oscillations with a shifted frequency
$\bar{\omega}^{2}=\omega^{2}-\frac{3}{16}\Lambda $,
or else (critical or sub-critical) exponential decay.
Notice that, in contrast with dark matter ripples, which are
only $1/t $-suppressed, dark energy ripples are exponentially
$\exp{(-\frac{1}{2}\sqrt{3\Lambda}t})$-suppressed.
This is expressed in Fig.\ref{a} where one can appreciate the
smooth dark matter to dark energy transition.
The cosmological $\Lambda$CDM-like evolution is recovered
(see the dashed lines in Fig.\ref{Hphi2}) by time averaging over
the ripples.
The total number $N$ of ripples at the matter dominated era
is estimated by
\begin{equation}
	N \simeq {\cal O} \left(
	\frac{\omega}{\pi\sqrt{3\Lambda}}\right) ~.
\end{equation}

The structure of the scalar potential $V(\phi)$ is quite arbitrary.
As long as a positive Newton constant VEV is our only physical
requirement, a quadratic potential will do.
It is then by no means trivial that the entire physical dark radiation
$\rightarrow$ matter $\rightarrow$ energy domination sequence 
is recovered at no extra cost.
But suppose we want more.
In particular, one would like to incorporate inflation\cite{inflation}
into the theory.
Traditionally, a scalar field is anyhow invoked for such a purpose,
so we are naturally led to a \emph{unified} inflationary model
where the inflaton and the reciprocal Newton 'constant' are in fact
the one and the same scalar field.
In which case, to end up with (say) a quartic Higgs-like
$V_{eff}(\phi)$, one better start, as dictated by eq.(\ref{Veff}), from
a cubic $V(\phi)$, namely
\begin{equation}
	V(\phi)=4\Lambda_{0}\phi(1-G\phi)^2
	+2G\Lambda\phi^{2} ~.
\end{equation}
The only constraint imposed on the cubic polynomial, namely
$V(0)=0$, which can be waived away in case of need, is engineered
to give rise to the symmetric (under $\phi\rightarrow -\phi$)
double-well effective potential
\begin{equation}
	V_{eff}(\phi)=
	\frac{\Lambda_{0}}{G^{2}}(1-G^{2}\phi^{2})^{2}~.
\end{equation}
The coefficient $\Lambda_{0}$ is to be identified as the
cosmological constant during the inflationary era.

\begin{figure}[ht]
	\includegraphics[scale=0.27]{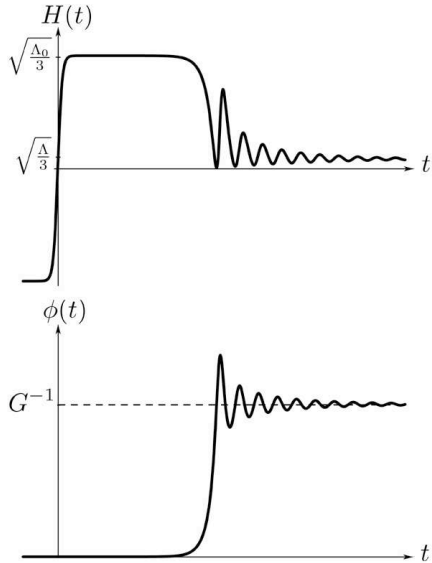}
	\caption{\label{inflation} The Hubble constant $H(t)$
	and the reciprocal Newton constant $\phi(t)$ evolution
	governed by a double-well $V_{eff}(\phi)$. The novel
	deflation $\rightarrow$ inflation transition is demonstrated.
	The after inflation evolution is the same as in Fig.\ref{Hphi2}.}
\end{figure}

\medskip
The non-physical (unstable) yet pedagogical solution $\phi(t)=0$,
that is sitting for ever on the top of the $V_{eff}(\phi)$ hill, is
unexpectedly correlated with a soliton-like solution for $H(t)$
\begin{equation}
	H(t)=\sqrt{\frac{\Lambda_{0}}{3}}
	\tanh\left(2\sqrt{\frac{\Lambda_{0}}{3}}t\right)  ~.
\end{equation}
It represents a smooth pre-Big-Bang\cite{pBB} transition
(borrowing string theory terminology, one may call it a 'graceful
exit'\cite{grace}) from eternal deflation to eternal inflation.
The general solution, however, introduces a constant of
integration $\chi>0$, such that around $\phi(0)=0$ we have
\begin{equation}
	\phi(t) \simeq \chi\sqrt{\frac{\Lambda_{0}}{3}}t ~.
\end{equation}
The so-called 'graceful exit' is now not only a deflation to
inflation transition, but furthermore a novel transition from
\emph{repulsive} ($\phi<0$) to \emph{attractive} ($\phi>0$)
gravity.
The duration of the inflationary period is of order
${\cal O}(-\sqrt{3/\Lambda_{0}}\log{\chi})$, and can be
arbitrarily fine tuned by adjusting $\chi$.
The full numerical solution, omitting the interesting
$\phi<0$ section which is a bit beyond the scope of this
Letter, is depicted in Fig.\ref{inflation}.
%%%%%%%  Fig.5

\medskip
One may find it conceptually difficult to go beyond GR, and accept
the unconventional idea that cosmological dark matter is in fact
nothing but a signature of a damped oscillating Newton constant.
After all, it is easier to phenomenologically parameterize
$\Lambda$CDM than struggle with the puzzle where does the
cosmological dark matter (and/or dark energy) actually come from.
In this paper, by minimally generalizing GR, Brans-Dicke style,
without explicitly introducing any additional matter fields or dust,
we have derived a cosmological evolution which serendipitously
resembles the dark radiation $\rightarrow$ dark matter $\rightarrow$
dark energy domination sequence.
The main prediction of our theory is that the FRW scale factor
associated with the emerging, necessarily general relativistic,
dark matter dominated Universe (see Fig.\ref{a}) is superimposed
with a rippled structure.
To be on the safe side, one can always seek shelter at the GR-limit
$\omega \rightarrow \infty$ where the ripples die away.
But on the other hand, one can only hope that the novel
mass scale $\omega$ is not that large, allowing the ripples to
be experimentally tested.
Our final remark concerns the time independent radially
symmetric solution which is still under active
investigation\cite{radial}.
At this stage, we can only confirm that the Newton limit has
been recovered, with the deviation from the $r^{-2}$ force law
being $e^{-\omega r}$ Yukawa suppressed.

\medskip
\begin{acknowledgments}
It is my pleasure to thank Professors Kameshwar Wali and Mark
Trodden for their cordial hospitality at Syracuse University.
Enlightening discussions with Professors Philip Mannheim,
Ray Volkas, and David Owen are gratefully acknowledged.
\end{acknowledgments}

\end{document}